\documentclass[conference]{IEEEtran}
\IEEEoverridecommandlockouts
\usepackage{cite}
\usepackage{amsmath,amssymb,amsfonts}
\usepackage{graphicx}
\usepackage{textcomp}
\usepackage{xcolor}
\usepackage{amsmath,amsfonts}
\usepackage{array}
\usepackage[caption=false,font=normalsize,labelfont=sf,textfont=sf]{subfig}
\usepackage{textcomp}
\usepackage{stfloats}
\usepackage{url}
\usepackage{verbatim}
\usepackage{graphicx}
\usepackage{cite}
\usepackage{graphicx}
\usepackage{algorithm}
\usepackage{algpseudocode}
\usepackage{amsmath}
\usepackage{booktabs} 
\usepackage{balance}

\def\BibTeX{{\rm B\kern-.05em{\sc i\kern-.025em b}\kern-.08em
    T\kern-.1667em\lower.7ex\hbox{E}\kern-.125emX}}
\begin{document}

\title{LLMs are All You Need? Improving Fuzz Testing for MOJO
with Large Language Models\\
}

\author{
    \IEEEauthorblockN{Linghan Huang\IEEEauthorrefmark{1}, 
                      Peizhou Zhao\IEEEauthorrefmark{1}, 
                      Huaming Chen\IEEEauthorrefmark{1}}
    \IEEEauthorblockA{\IEEEauthorrefmark{1}School of Electrical and Computer Engineering, University of Sydney, Sydney, Australia\\
    Email: \{lhua5130, pzha2332\}@uni.sydney.edu.au, huaming.chen@sydney.edu.au}
}


\maketitle

\begin{abstract}
    The rapid development of large language models (LLMs) has revolutionized software testing, particularly fuzz testing, by automating the generation of diverse and effective test inputs. This advancement holds great promise for improving software reliability. Meanwhile, the introduction of MOJO, a high-performance AI programming language blending Python's usability with the efficiency of C and C++, presents new opportunities to enhance AI model scalability and programmability. However, as a new language, MOJO lacks comprehensive testing frameworks and a sufficient corpus for LLM-based testing, which exacerbates model hallucination. In this case, LLMs will generate syntactically valid but semantically incorrect code, significantly reducing the effectiveness of fuzz testing. To address this challenge, we propose MOJOFuzzer, the first adaptive LLM-based fuzzing framework designed for zero-shot learning environments of emerging programming languages. MOJOFuzzer integrates a mutil-phase framework that systematically eliminates low-quality generated inputs before execution, significantly improving test case validity. Furthermore, MOJOFuzzer dynamically adapts LLM prompts based on runtime feedback for test case mutation, enabling an iterative learning process that continuously enhances fuzzing efficiency and bug detection performance. Our experimental results demonstrate that MOJOFuzzer significantly enhances test validity, API coverage, and bug detection performance, outperforming traditional fuzz testing and state-of-the-art LLM-based fuzzing approaches. Using MOJOFuzzer, we have conducted a first large-scale fuzz testing evaluation of MOJO, uncorvering 13 previous unknown bugs. This study not only advances the field of LLM-driven software testing but also establishes a foundational methodology for leveraging LLMs in the testing of emerging programming languages.

\end{abstract}


\maketitle

\section{Introduction}

The rapid advancement of artificial intelligence (AI) across various critical domains, such as autonomous driving systems, medical diagnostics, and financial services, has led to its widespread adoption and substantial growth. However, developing and deploying AI in production environments presents significant engineering challenges. To address the inherent complexities associated with machine learning and AI development, novel infrastructure and programming language solutions have emerged. A notable example is the \textbf{MOJO} programming language, developed by the Modular team led by Chris Lattner, who creates LLVM~\cite{LLVM:CGO04} and Swift~\cite{apple_2019}. MOJO blends Python's syntactic simplicity and flexibility with the efficiency and low-level performance characteristics of system languages like C and C++. Recent benchmark has shown that MOJO can achieve speedups of up to 68,000 times compared to the current version of Python~\cite{awan_2023}. It significantly bridge the gap between research prototyping and production AI deployment.

By employing a progressive migration strategy, MOJO is tightly integrated with CPython, allowing existing Python code to be executed within the MOJO compilation environment without changes. It also enables developers to write highly portable and efficient code that rivals the performance of C while maintaining seamless compatibility with Python ecosystem. This strategic design facilitates a smooth transition for developers and enhances the accessibility of high-performance computing in AI workflows. Furthermore, MOJO incorporates advanced system programming and meta-programming features, making it a forward-thinking solution for next-generation programming paradigms~\cite{thomason_2024}. Ultimately, MOJO represents an important step towards enhancing AI-driven application's efficiency and computational performance, fostering the broader adoption and accelerated progress of AI technologies. As an emerging programming language designed explicitly for AI, MOJO has gained significant attention for its performance and emphasis on secure development. Meanwhile, the emergence of large language models (LLMs) has transformed software testing, particularly fuzz testing, by enhancing its depth and efficiency. Recent works, such as TitanFuzz~\cite{deng2023large} and FuzzGPT~\cite{deng2024large}, demonstrate how integrating LLMs into fuzzing can greatly improve the bug discovery and testing effectiveness. However, current LLM-based fuzzers typically require extensive training corpora specific to the system under test (SUT). It limits their applicability to novel or emerging programming language like MOJO. 

To address this challenge, we introduce MOJOFuzzer, the first LLM-based fuzz testing framework specifically designed for the MOJO language. MOJOFuzzer strategically integrates two distinct approaches to achieve optimal performance in generating novel and effective test cases: (1) leveraging LLM-driven fuzzing to automatically generate diverse, novel and effective test cases; (2) fine-tuning open-source LLMs with curated datasets, including historical bug patterns and language-specific syntax, enhancing the semantic validity of generated inputs. A key challenge faced by LLM-driven fuzzers is \textit{model hallucination}, where models will generate syntactically valid but semantically incorrect or ineffective test cases. We observe that this is particularly exacerbated in zero-shot environments due to the absence of domain-specific context. MOJOFuzzer address hallucination by incorporating targeted fine-tuning and structured adaptive prompt engineering, guiding LLMs to align closely with MOJO's syntax and semantics. Our experimental results demonstrate MOJOFuzzer significantly increase the test case validity and effectiveness. In summary, our work makes the following contributions:
\begin{itemize}
    \item \textbf{Effective mitigation of LLMs hallucinations in zero-shot fuzzing.} 
    MOJOFuzzer is the first LLM-driven fuzz testing framework designed to address the challenges in \textit{zero-shot learning environments}, where SUT is absent from existing model corpora. MOJOFuzzer is specifically tailored for MOJO with a novel integration of fine-tuning and prompt engineering techniques, which substantially reduce hallucinations. Our method ensures that generated test inputs reliably trigger meaningful behaviors and software defects, even without prior domain-specific data.
    \item \textbf{Real-world bug discovery and security implications.} Using MOJOFuzzer, we have successfully uncovered 13 unknown bugs, including potential security vulnerabilities. Of these, 9 bugs have been confirmed and patched by the MOJO team. Our empirical evaluation demonstrates that MOJOFuzzer significantly outperforms existing fuzzing techniques and state-of-the-art LLM-based fuzzers, achieving superior mutation efficiency, API coverage, and defect detection rates.
\end{itemize}

\section{RESEARCH QUESTIONS \& MOTIVATION}
In this section, we discuss the design of our research questions, together with the motivation.
\subsection{RQ1: How effective is MOJOFuzzer in generating valid and diverse test inputs for the MOJO language in a zero-shot learning environment compared to existing fuzzing tools?}
Recent studies indicate that LLM-based fuzzers generally outperform traditional fuzzing methods by generating more diverse and effective test seeds~\cite{8863940}. However, existing LLM-based fuzzers heavily rely on extensive pre-existing training data~\cite{wang2024softwaretestinglargelanguage}, limiting their effectiveness in zero-shot environments like for MOJO, which lack such resources. This limitation significantly increases the risk of model hallucination, where LLMs generate syntactically plausible, yet semantically invalid test cases, thereby reducing testing efficacy~\cite{9832795}. To bridge this gap, MOJOFuzzer leverages a novel and strategical approach of fine-tuning with minimal domain-specific data of MOJO syntax and historical bug fragments, and adaptive prompt engineering to mitigate the hallucination issue and improving seed quality. We aim to empirically verify whether MOJOFuzzer significantly improves the validity, diversity, and bug-detection capabilities of generated test inputs compared to both traditional and existing LLM-based fuzzing methods.\\ 
\subsection{RQ2: What are the key components that enables MOJOFuzzer to effective reduce LLM hallucination and improve the efficiency and validity of fuzz-generated test cases?}
Existing LLM-based fuzzers typically adopt either fine-tuning strategies, which requires substantial training resources~\cite{meng2024large}, or general-purpose LLMs with minimal customization, which increases the risk of hallucinations~\cite{wang2024softwaretestinglargelanguage}. In contrast, MOJOFuzzer adopts an adaptive, runtime-aware prompt refinement approach combined with lightweight fine-tuning specifically tailored to MOJO. This design effectively overcome the hallucination challenge without extensive domain-specific training data. Through well-design controlled experiments, we systematically analyze the impact of each module in MOJOFuzzer to reveal how these factors collectively contribute to MOJOFuzzer’s superior performance.\\

\subsection{RQ3: Can MOJOFuzzer uncover complex or previously unknown bugs in real-world scenarios for MOJO, and what insights does this provide about the strengths and limitations of LLM-based zero-shot fuzzing in practice?}
Due to the novelty of the MOJO language, its current systematic testing is limited, yet systematic bug testing remains largely unexplored. We observe that existing fuzzers may struggle to identify complex or subtle bugs due to challenges in handling zero-shot contexts and LLM hallucinations~\cite{jiang2024fuzzingmeetsllmschallenges}. To evaluate MOJOFuzzer's practical effectiveness, we have conducted comprehensive fuzz testing on MOJO, aiming to identify previously unknown bugs and security vulnerabilities. For example, our approach flags any input that triggers a compilation or runtime error as a potential defect, which we then manually inspect to confirm true positives. To date, MOJOFuzzer has uncovered 13 critical bugs, of which 9 bugs have been confirmed and fixed by MOJO community. These results  highlights both the strengths and limitations of the zero-shot LLM-based fuzzing technique and the MOJO language, providing concrete insights for future research and tool development. 

\section{BACKGROUND}
\subsection{MOJO Language}
MOJO is a modern programming language developed by Modular AI, specifically designed for AI. It integrates key features from Python, C++, and Rust, combining Python's expressive syntax with the computational efficiency of C++ and Rust. While Python is widely recognized for its simplicity and ease of use, C++ and Rust provide superior performance and system-level control. MOJO retains Python-like syntax while achieving significantly higher execution speeds. According to~\cite{mohan2023}, a key feature of MOJO is its direct integration with most Python libraries, allowing legacy Python code to run seamlessly within MOJO without modification. Furthermore, MOJO employs Multi-Level Intermediate Representation (MLIR)~\cite{mlir2024}, enabling developers to optimize AI workloads through efficient hardware acceleration and vectorized computations, thereby enhancing performance and mitigating concurrency-related issues. Additionally, MOJO supports \textit{zero-cost abstraction}, granting developers fine-grained control over memory allocation while maintaining high execution efficiency. These features make MOJO particularly suitable for AI development. However, as a newly developed language, MOJO currently lacks mature testing frameworks, which increases the likelihood of undiscovered software bugs and security vulnerabilities.\\ 

\subsection{LLM-based Fuzzer}
Software bugs and defects can result in unpredictable behavior or even catastrophic failures in software systems. Traditional testing methods, such as unit testing~\cite{article} and integration testing~\cite{131377}, validate the functionality of individual system components. However, these approaches often fail to detect deeper systemic issues~\cite{wang2024softwaretestinglargelanguage}. The introduction of fuzz testing addresses these limitations by systematically injecting a large number of random test inputs and monitoring the system outputs~\cite{10.1145/3243734.3243804}. This approach enables the discovery of latent vulnerabilities and facilitates comprehensive stability assessments. Recently, fuzz testing has been largely enhanced with the integration of Large Language Models (LLMs)~\cite{zhao2023surveylargelanguagemodels}, which are trained on extensive Internet text corpora. LLMs have demonstrated exceptional performance in various natural language processing tasks. LLMs are based on the Transformer architecture\cite{vaswani2023attentionneed} and are generally categorized into Encoder-only~\cite{devlin2019bert}, Decoder-only~\cite{fu2023decoderonlyencoderdecoderinterpretinglanguage}, and Encoder-Decoder~\cite{fu2023decoderonly} models. Developers can employ prompt engineering to optimize these models for specific tasks~\cite{chen2023unleashing}; alternatively, LLMs can be fine-tuned with domain-specific data. 

\begin{figure}[h]
    \centering
    \includegraphics[width=.8\columnwidth]{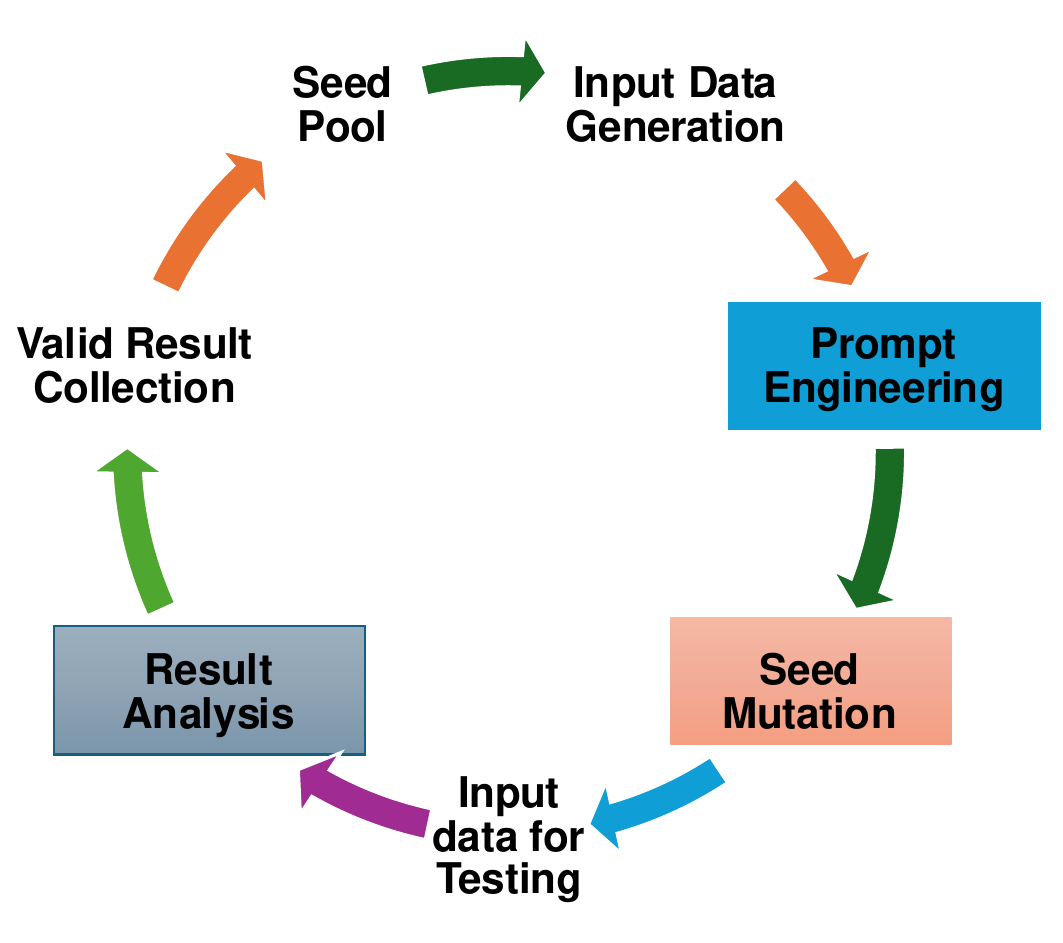}
    \caption{An overview of LLM-based Fuzzer}
    \label{fig:llm}
\end{figure}

While fine-tuned LLMs~\cite{fu2022effectiveness} have been increasingly utilized to tackle complex code analysis and software testing challenges, it has brought more possibilities for fuzz testing. LLMs can systematically address distinct testing objectives by leveraging extensive training data and structured prompt engineering. Recent studies on LLM-based fuzzers typically follow a general workflow, as illustrated in Figure\ref{fig:llm}~\cite{wang2024softwaretestinglargelanguage,huang2024largelanguagemodelsbased}, integrating LLMs at various stages of fuzz testing to enhance seed diversity, input validity, and bug discovery capability.

\subsubsection{Fuzzer by LLMs}
In Figure 1, integrating LLMs into the seed mutation process of traditional fuzzers, along with strategic prompt engineering techniques, significantly improve test seed diversity and effectiveness~\cite{deng2023large,10.1145/3605157.3605173,xia2024fuzz4alluniversalfuzzinglarge, yang2023whitebox, meng2024large, wu2023smart, hu2023augmentinggreyboxfuzzinggenerative, liu2023testing, qiuchemfuzz, googleblogAIPoweredFuzzing}. As a representative LLMs based fuzzer, TitanFuzz leverages LLMs to generate initial input seeds by providing specific prompts to the models. The LLM then automatically mutates these seeds through iterative refinement, after which they are used to test deep learning libraries. Similarly, ChatAFL employs an LLM to first learn the grammar of a given protocol, thereby facilitating a structure-aware mutation~\cite{meng2024large}. This approach enhances seed diversity, thereby improving the efficiency of network protocol testing. As for Fuzz4All, it targets multiple programming languages, including Python, Java, and C~\cite{xia2024fuzz4alluniversalfuzzinglarge}. In this approach, an LLM processes distilled user inputs as prompts to generate initial test seeds. At the end of each iteration, the LLM assists the fuzzer by executing various seed mutation strategies, thereby improving the effectiveness of the tests. Additionally, ParaFuzz exploits LLMs interpretability to refine model predictions via sophisticated prompt engineering, significantly improving defect detection in NLP models\cite{yan2023parafuzzinterpretabilitydriventechniquedetecting}.


\subsubsection{Fine-tuning fuzzers}
Different from LLM-based fuzzers (\textit{Fuzzer by LLMs}) that rely on general pre-trained models without fine-tuning~\cite{yan2023parafuzzinterpretabilitydriventechniquedetecting,xia2024fuzz4alluniversalfuzzinglarge,meng2024large,zhang2023understanding,jha2023bertrlfuzzer,jin2024llmsllmbasedagentssoftware}, fine-tuning fuzzers leverage LLMs that have been specifically fine-tuned on domain-specific datasets. One most recent example is FuzzGPT, a framework specifically designed for fuzz testing deep learning libraries~\cite{deng2024large}. In particular, the effectiveness of such fine-tuned fuzzers is heavily dependent on the capabilities of their underlying LLMs. FuzzGPT integrates multiple high-performance LLMs, including CodeX~\cite{finnie2022robots}, InCoder~\cite{fried2023incodergenerativemodelcode}, and CodeGen~\cite{nijkamp2022codegen}, as key components of its fuzzing framework. This approach involves fine-tuning these LLMs on a substantial dataset of historical bug-triggering code snippets to refine their model parameters. Once the LLMs have internalized these bug patterns, it is expected that high-quality test seeds will be generated either by modifying historical bug-triggering code or by synthesizing entirely new test cases reflecting those patterns. As illustrated in Figure.~\ref{fig:llm}, these LLM-based fuzzers primarily utilize LLMs for seed generation and mutation. The efficacy of LLMs in these tasks largely stems from both their ability to leverage extensive high-quality and domain specific training data, as well as the pre-existing knowledge they have acquired about the target systems, which are typically well-established. An LLM trained on a large volume of domain-specific data has been shown to generate fuzzing inputs that surpass those produced by traditional fuzzers~\cite{wang2024softwaretestinglargelanguage}. However, it will be significant impacted when dealing with the MOJO programming language, where such data is scarce or nonexistent. In this paper, we investigate how zero-shot LLM-based fuzzing, without relying on extensive training corpora, can address these limitations and effectively uncover software defects in novel programming languages such as MOJO.


    

\section{The OVERVIEW of MOJOFuzzer}

\begin{figure*}[ht]
\centering
\includegraphics[width=\textwidth]{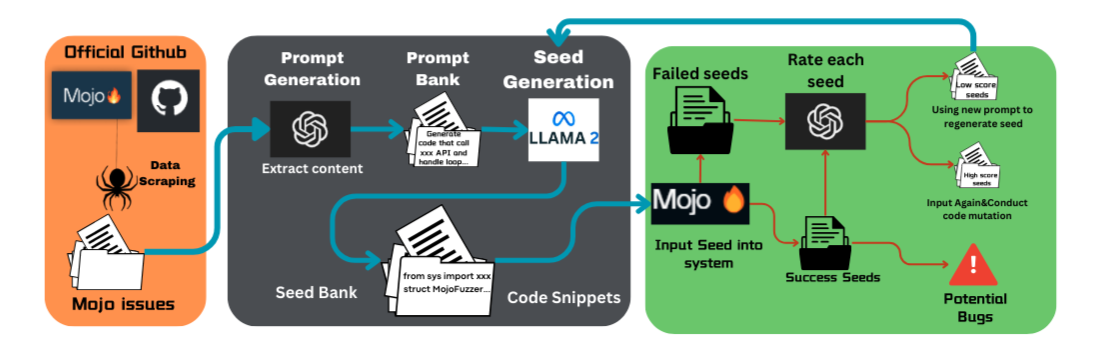} 
\caption{Overview of MOJOFuzzer}
\label{fig:mojofuzzer}
\end{figure*}

Figure~\ref{fig:mojofuzzer} illustrates the detailed architectural overview of the proposed MOJOFuzzer framework. The overall framework begins with the automated collection of related datasets, followed by a preliminary data cleaning and preprocessing step to ensure data quality and consistency (Section.~\ref{Sec:datapre}). Subsequently, an initial prompt bank will be generated using LLMs with crafted prompts. Leveraging this prompt bank, a fine-tuned LLM (Section.~\ref{Sec:llmfinetuning}), specifically trained to align with MOJO syntax and domain-specific semantics, will create corresponding test seeds, thereby populating the initial seed bank (Section.~\ref{Sec:prombank}). These seeds are employed in an initial round of fuzz testing on MOJO (Section.~\ref{Sec:ini}). 

To further enhance the quality and diversity of generated test seeds, we introduce a novel mutation algorithm to maximize the exploration of input space (Section.~\ref{Sec:mut}). The mutation strategy incorporates a mutation-scoring mechanism, assigning each seed a quantitative measure of its observed effectiveness and impact during fuzzing execution. According to the mutation score, test seeds are classified into two distinct categories for following mutation approach. Seeds with higher mutation scores undergo a half-mutation process (Section.~\ref{Sec:half}), involving direct mutations to existing test cases. Conversely, seeds with lower scores are subjected to full mutation (Section.~\ref{Sec:full}), where the test inputs will be regenerated from their initial prompts to facilitate broader and more diverse operations. This mutation-scoring mechanism enables MOJOFuzzer to effectively prioritize valuable seeds and efficiently directs mutation resources towards designated test seeds. Ultimately, MOJOFuzzer significantly optimizes the overall fuzz testing efficiency, maximizing bug-detection capability for the MOJO language.


\subsection{Initiation}\label{Sec:ini} 
\subsubsection{Dataset Preparation}\label{Sec:datapre}
To create a reliable corpus for MOJO fuzz testing, we systematically collected relevant data from public sources, including the official MOJO GitHub repository and related online documentation~\cite{githubGitHubModularmax,modularMojoManual}. The data extraction process primarily focuses on three types of resources:

\begin{itemize}
    \item \textbf{Bug reports and crash logs} associated with the MOJO language, reported on the official GitHub repository;
    \item \textbf{Syntax rules and language constructs} for MOJO language, including representative code snippets and structural definitions;
    \item \textbf{Official documentation} including  API specifications, user manual and function reference guide.
\end{itemize}

Upon collecting these datasets, a rigorous preprocessing was performed to ensure data quality and relevance. This step includes following key points:
\begin{itemize}
    \item Filtering out bug reports or error logs that were unrelated to MOJO to ensure dataset relevance;
    \item Eliminating incomplete, inconsistent, or ambiguous syntax descriptions and code snippets to maintain data integrity;
    \item Standardizing the user manual and API guide by removing redundant information and retaining only core technical content.
\end{itemize}

Following this rigorous pre-processing, we successfully curated a refined dataset comprising approximately $\sim$300 unique bug reports and $\sim$1,500 distinct data samples containing MOJO syntax rules, examles, and language documentation. This dataset was subsequently employed to fine-tune the LLMs, which we have opted in LLAMA2 13B language model. It significantly enhances the model's capability to generate syntactically accurate and semantically meaningful test seeds specific for MOJO.

\subsubsection{Prompt bank and Seed bank}\label{Sec:prombank}
The integration of prompt banks and seed banks has been widely recognized in recent research as an essential mechanism to systematically store and manage test seeds in LLM-driven fuzz testing frameworks~\cite{10.1145/3460319.3464795}. Within the MOJOFuzzer framework, these two components are essential for optimizing the fuzz testing workflow, thus significantly enhancing fuzzing efficiency and coverage. Since the effectiveness of the fuzzing process heavily relied on the quality and diversity of initial seed samples, MOJOFuzzer utilizes structured instructions to guide model output during initial seed generation. These prompts provide crucial contextual guidance, ensuring that generated test cases are syntactically valid, semantically relevant, and diverse to the fuzzing objectives. Prompt Bank serves as a repository for structured prompt templates, facilitating the generation of initial and mutated seeds. To optimize computational resource usage and maintain scalability, MOJOFuzzer adopts advanced prompting techniques, specifically Chain of Thought (CoT) prompting~\cite{wei2023chainofthoughtpromptingelicitsreasoning} and Role Prompting~\cite{kong2024betterzeroshotreasoningroleplay}. These techniques allow for structured decomposition of complex prompts, substantially improving model interpretability and adaptability. Empirical evaluations indicate that this approach reduces prompt text volume by approximately 30\% while preserving its semantic integrity and improving overall seed quality. The key components of the prompts are designed as follows. 
\begin{enumerate}
    \item \textbf{Syntax Analysis:}  
    `Please analyze the MOJO syntax in detail as the basis for processing data.'
    
    \item \textbf{Role-Based Framing:}  
    `As an analyst, you will extract core questions and related codes from complex data.'
    
    \item \textbf{Automated Data Filtering:}  
    `Extract and retain functionally relevant code while removing redundant metadata. Prioritize recursive functions and complex control structures.'
    
    \item \textbf{Content Summarization:}  
     `Summarize each problem into the most critical description, including a brief description of the possible MOJO language problems, specific code reproduction steps, and the execution of the code in a specific version environment.'
    
    \item \textbf{Prompt seeds Generation:}  
    `Generate prompt bank seeds based on simplified and cleaned data. Make sure each seed focuses on showing how to cause a specific problem with the MOJO language, including version information and reproduction steps.'
\end{enumerate}

\subsubsection{Seed Bank}
The \textbf{Seed Bank} within MOJOFuzzer systematically stores and manages test seeds generated from the structured prompts provided by the Prompt Bank. The repository maintains both original and mutated test seeds that evolve through subsequent fuzzing iterations, ensuring an organized, traceable, and efficient fuzz testing process. By systematically leveraging these test seeds, MOJOFuzzer effectively explores diverse execution paths and potential software vulnerablities within MOJO language. For example, when testing the \texttt{def} function type in MOJO, each prompt in the Prompt Bank explicitly guides seed generation and mutation through clearly defined components, which are then used as input to evaluate the system's behavior. 

\begin{enumerate}
    \item \textbf{Problem Description:} `The identified issue in the MOJO language pertains to the following potential bugs: [code reproduction description]. These issues may impact system performance, cause compilation failures, or introduce security vulnerabilities, necessitating rigorous testing.'
    
    \item \textbf{Test Objective:} `Generate code snippets that evaluate how MOJO processes [brief description]. The generated tests should validate whether the system correctly handles expected behaviors and reports errors when necessary.'
    
    \item \textbf{Test Case Generation:} `Based on the problem description, create test cases that can trigger relevant bugs. For example, if the issue involves data type conversions, generate code snippets that test a range of boundary values, invalid type coercions, and unexpected input structures.'
    
    \item \textbf{System Environment Specification:} `All test cases should be executed within a defined environment, specifying details such as the operating system, MOJO compiler version, library dependencies, and runtime configurations.'
    
    \item \textbf{Execution and Verification:} `Each generated test case should be executed, and its results systematically validated. System responses, including error messages, execution failures, program crashes, or unexpected behavior, must be recorded and reported for further analysis.'
\end{enumerate}

Using such structured prompts, MOJOFuzzer efficiently generates high-quality initial seeds tailored for MOJO language constructs. To facilitate dynamic seed management and retrieval during fuzzing, we establish an indexing mechanism for `Prompt Seeds' and `Code Seeds'. As depicted in Algorithm~\ref{alg:1}, this association enables MOJOFuzzer to iterative adapt test inputs based on real-time testing feedback, which significantly enhances the diversity, efficiency, and effectiveness of MOJO language testing, ensuring systematic and comprehensive defect discovery.

\begin{algorithm}
\footnotesize
\caption{Seed Generation Process}
\begin{algorithmic}[1]
\Function{SeedGenerationProcess}{PromptsBank}
    \State \textbf{Output:} SeedsBank
    \State SeedsBank $\gets$ create\_empty\_SeedsBank()
    \State prompts $\gets$ PromptsBank.get\_all\_prompts()
    \For{each prompt in prompts}
        \State generated\_seed $\gets$ seedGeneration(prompt)
        \State SeedsBank.add\_seed(generated\_seed)
    \EndFor
    \State \Return SeedsBank
\EndFunction
\end{algorithmic}
\label{alg:1}
\end{algorithm}

 \subsection{Mutation}\label{Sec:mut}
\subsubsection{Mutation Strategy}\label{Sec:mutstr}

Algorithm~\ref{alg:2} outlines the mutation strategy employed by MOJOFuzzer. Upon completion of the initial seed generation phase, the framework systematically aligns the datasets stored in the Prompt Bank and Seed Bank to facilitate subsequent mutation operations. During execution by MOJO compiler, the runtime behavior will be automatically captured, including successful execution results, failures, crashes, and any anomalous behaviors for further analysis. These results will inform the analysis process prior to the next step. The test seeds will be classified based on their observed effectiveness and impacts. Specifically, seeds that either trigger unexpected behavior or uncover potential software defects (i.e., `successful seeds') are assigned a higher mutation priority due to their demonstrated ability to effectively explore sensitive execution paths. Seeds that fail to produce meaningful outcomes are also retained but assigned a lower priority. This prioritization strategy ensures focused exploration around critical test scenarios, maximizing fuzzing effectiveness. 

To systematically quantify the mutation workflow, we introduce a mutation scoring mechanism. Seeds assigned higher mutation scores enter the Half Mutation procedure, while those with lower scores undergo Full Mutation. The mutation score ($S_{\text{mutation}}$) quantitatively assesses whether a seed requires further processing. High-scoring seeds are directly subjected to code-level mutations to preserve their underlying structural patterns, whereas low-scoring seeds trigger modifications at the prompt level, subsequently regenerating code through the fine-tuned model. Formally, the mutation score is calculated as follows:

\[
S_{\text{mutation}} = N_{\text{API}} + C_{\text{complexity}}
\]
\begin{itemize}
    \item \textbf{Number of API calls ($N_{\text{API}}$):} We calculate the number of API calls for each code seed. Test seeds with higher call counts tend to cover a wider range of MOJO language functions to get higher scores.
    \item \textbf{Code complexity ($C_{\text{complexity}}$):} We evaluate the code complexity of the seeds, including the nesting depth of functions, control flow structure, and code length. Seeds with higher complexity tend to trigger more boundary conditions to get higher scores~\cite{arora_2017}.\newline
    $C_{\text{complexity}}$ follow the rules:
        \begin{itemize}
            \item $C_{\text{complexity}} = score 1$ when $O(1)$;
            \item $C_{\text{complexity}} = score 2$ when $O(n)$;
            \item $C_{\text{complexity}} = score 3$ when $O(n \log n)$;
            \item $C_{\text{complexity}} = score 4$ when $O(n^2)$;
            \item $C_{\text{complexity}} = score 5$ when $O(n^3)$ or more complex.
        \end{itemize}
\end{itemize}

\begin{algorithm}
\caption{Fuzzing with Seeds}
\scriptsize
\begin{algorithmic}[2]
\Function{FuzzingWithSeeds}{PromptsBank, SeedsBank}
    \State \textbf{Output:} SuccessSeeds
    \State SuccessSeeds $\gets$ create\_empty\_SuccessSeedsBank()
    \State maxIterations $\gets$ 10, iteration $\gets$ 0
    \While{(PromptsBank.is\_not\_empty() or SeedsBank.is\_not\_empty()) and iteration < maxIterations}
        \State iteration $\gets$ iteration + 1
        \State currentPrompt $\gets$ PromptsBank.get\_next\_prompt()
        \State currentSeed $\gets$ SeedsBank.get\_next\_seed()
        
        \If{FuzzingTest(currentSeed, target="mojo programming language")}
            \State SuccessSeeds.add(currentPrompt, currentSeed)

            \State score $\gets$ ComputeScore(currentSeed)
            
            \If{score > 50}
                \State register\_as\_potential\_bug(currentPrompt, currentSeed)
            \Else
                \State mutatedSeed $\gets$ SeedMutation(currentSeed)
                \If{FuzzingTest(mutatedSeed, target="mojo programming language")}
                    \State SuccessSeeds.add(currentPrompt, mutatedSeed)
                \Else
                    \State SeedsBank.add\_seed(mutatedSeed)
                \EndIf
            \EndIf

        \Else
            \State score $\gets$ ComputeScore(currentSeed)
            \If{score <= 50}
                \State mutatedSeed $\gets$ SeedMutation(currentSeed)
                \If{FuzzingTest(mutatedSeed, target="mojo programming language")}
                    \State SuccessSeeds.add(currentPrompt, mutatedSeed)
                \Else
                    \State SeedsBank.add\_seed(mutatedSeed)
                \EndIf
            \Else  
                \State mutatedPrompt $\gets$ PromptMutation(currentPrompt)
                \State newSeed $\gets$ seedGeneration(mutatedPrompt)
                \State SeedsBank.update\_seed(currentSeed, newSeed)
                \If{FuzzingTest(newSeed, target="mojo programming language")}
                    \State SuccessSeeds.add(mutatedPrompt, newSeed)
                \Else
                    \State SeedsBank.add\_seed(newSeed)
                \EndIf
            \EndIf
        \EndIf
        \State \textbf{print}("Iteration:", iteration, "completed.")
    \EndWhile
    \State PromptsBank.clear()
    \State SeedsBank.clear()
    \State \Return SuccessSeeds
\EndFunction

\end{algorithmic}
\label{alg:2}
\end{algorithm}


\subsubsection{Half Mutation}\label{Sec:half}
The primary objective of Half Mutation is to perform targeted mutations on test seeds. A mutation strategy will be randomly selected from a predefined library, which is then encoded as a structured textual prompt and combined with the original seed code snippet. The result will be provided as input to the fine-tuned LLAMA2 model for mutation. Following we discuss these mutation strategies and their corresponding prompt examples:
\begin{enumerate}
    \item \textbf{Strategy 1: Method Augmentation}\newline \textbf{Objective:} Introduce additional methods to expand the execution paths of existing code without altering its original logic.\newline \textbf{Prompt Example:} \textit{`Please augment the following code snippet by adding new methods. Ensure the original functional logic remains unchanged.'}
    
    \item \textbf{Strategy 2: Method and Parameter Substitution}\newline
    \textbf{Objective:} Replace existing methods or parameters with alternative implementations while preserving structural integrity and overall functionality.\newline
    \textbf{Prompt Example:} \textit{`Please substitute existing methods and parameters in the following code with logically equivalent alternatives. The modified code should retain its original functionality and logical consistency.'}
    
    \item \textbf{Strategy 3: Code Optimization and Refinement}\newline
    \textbf{Objective:} Refine code structure by rearranging statements and optimizing algorithms without modifying the core logical behavior.\newline
    \textbf{Prompt Example:} \textit{`Optimize the structure and algorithmic efficiency of the following code snippet without changing its core logic. The goal is to enhance performance and maintain original functionality.'}
    
    \item \textbf{Strategy 4: Comprehensive Mutation}\newline
    \textbf{Objective:} Integrate multiple mutation strategies simultaneously to assess the code's stability and reliability under various changes.\newline
    \textbf{Prompt Example:} \textit{`Perform a comprehensive mutation of the following code by combining method augmentation, parameter substitution, and structural optimization strategies. Each mutation should preserve the original logical integrity while evaluating the robustness of the implementation.'}

\end{enumerate}

\subsubsection{Full Mutation}\label{Sec:full}
The fundamental principle underlying the Full Mutation strategy is to perform mutations at the prompt level, specifically targeting the prompt seeds derived from previous iterations (including both prompt and code seeds). After mutation, the modified prompt seeds are subsequently reintroduced into the model to generate new set of code seeds. We propose the following prompt mutation strategies to systematically guide the model's mutation process:

\begin{enumerate}
    \item \textbf{Strategy 1: Keyword Replacement}\newline \textbf{Description:} Identify critical keywords or phrases within the original prompt and substitute them with semantically equivalent or related terms.\newline \textbf{Example:} Original prompt \textit{`Generate a sorting algorithm'} is mutated into \textit{`Write a permutation function'}.

    \item \textbf{Strategy 2: Constraint Augmentation}\newline
    \textbf{Description:} Introduce specific constraints or detailed conditions within the prompt to narrow the scope of the model's generation space, thereby guiding more targeted code generation.\newline
    \textbf{Example:} Original prompt \textit{`Generate a sorting algorithm'} is mutated into \textit{`Generate a sorting algorithm implemented using recursion'}.
    
    \item \textbf{Strategy 3: Introduction of Defect Patterns}\newline
    \textbf{Description:} Explicitly incorporate common software defects or erroneous logic patterns into the prompts, encouraging the model to produce code seeds capable of triggering or uncovering specific errors.\newline
    \textbf{Example:} Original prompt \textit{`Generate a sorting algorithm'} is mutated into \textit{`Generate a sorting algorithm containing boundary condition errors'}.
    
    \item \textbf{Strategy 4: Composite Prompt Mutation}\newline
    \textbf{Description:} Apply multiple mutation strategies simultaneously to create prompt seeds that integrate various mutations, enhancing the robustness and effectiveness of the generated code seeds.\newline
    \textbf{Implementation:} A single prompt is simultaneously subjected to multiple transformations, such as keyword replacement, constraint addition, and defect pattern introduction, to evaluate code stability comprehensively.
\end{enumerate}
Unlike Half Mutation which directly mutates code-level seeds, Full Mutation retrieves corresponding prompts from Prompt Bank and initiates mutation directly at the prompt level. Existing literature focuses mainly on direct mutation at code snippet~\cite{huang2024largelanguagemodelsbased}. In contrast, within an LLM-driven fuzzing framework, we observe that prompt-level modifications significantly impact the structure and quality of the generated code seeds. Consequently, rather than conducting traditional direct code mutations commonly found in conventional fuzz testing methodologies~\cite{291011}, our approach also emphasizes systematic prompt altering. Performing mutations at the prompt level fundamentally alters the generation behavior and structural properties of code seeds, thereby introducing greater diversity and potentially achieving superior coverage in fuzz testing~\cite{sahoo2024systematicsurveypromptengineering}.

\subsubsection{Fine-tuning}\label{Sec:llmfinetuning}
Inspired by the approach in~\cite{deng2024large}, we further fine-tune LLMs for MOJO, in which we have selected LLAMA2 13B model~\cite{meta_2024}. The fine-tuning process utilizes two types of datasets: one capturing the syntactic rules and grammatical constructs of MOJO, and the other consisting of historical bug records. The goal is to improve the test coverage and code generation capabilities of the MOJOFuzzer framework. 
In the first stage, we train the model using MOJO grammar dataset. Each training sample includes an API name, a grammar description, and a corresponding code snippet. This training allows the model to learn correct grammatical structures, API patterns and semantic nuances of MOJO, ensuring that the generated code seeds adhere to the fundamental specifications of MOJO. 
Thus, the first stage focuses on enhancing the model's understanding of MOJO syntax and ensuring that its output is grammatically accurate. 
In the second stage, we fine-tune the model on a dataset of historical bugs. Each training sample provides a bug description, a triggering API, and an associated code snippet, encapsulating code patterns that led to known bugs. By capturing these defect patterns, the model learns to generate seeds specifically designed to uncover subtle, complex or previously unknown software defects. As a result, the second stage equips the model to not only produce specification-compliant code but also proactively generate seeds that expose system bugs during fuzz testing.

\section{RESULT ANALYSIS}
\subsection{Experimental Setup \& Evaluation Metrics}\label{Sec:met}
We detail the experimental design, including data collection procedure, utilized LLMs, training strategies, hardware infrastructure and evaluation metrics. To construct the dataset, we developed a specialized web crawler to systematically collect relevant data from GitHub repositories and official MOJO documentation~\cite{modularMojoManual}. Specifically, we obtained approximately 1,800 data samples, consisting of 300 historical bug reports and around 1,500 entries detailing MOJO syntax rules and usage guides. Subsequently, manual data cleaning procedures were applied to ensure the accuracy and relevance of collected data for fine-tuning purposes. We employed the GPT API provided by OpenAI~\cite{openai_2024} as a general-purpose model to generate prompt templates. For code seed generation and mutation, we fine-tuned the LLAMA2 13B model from Hugging Face, adopting the low-rank adaptation (LoRA) technique~\cite{hu2021loralowrankadaptationlarge} to optimize computational efficiency. During fine-tuning, the first stage involved learning syntactic structures and semantic rules of MOJO from grammar-focused datasets, while the second stage leveraged historical bug records to enhance the model's capability in detecting potential defects. All the experiments were conducted on a hardware platform with NVIDIA A6000 ada, which ensured sufficient computational resources for model training and fuzz testing tasks. We have included the state-of-the-art traditional and LLM-based fuzzers for evaluation, including MojoCoder~\cite{raihan2024mojobenchlanguagemodelingbenchmarks}, Fuzz4All~\cite{xia2024fuzz4alluniversalfuzzinglarge}, TitanFuzz~\cite{deng2023large}, and GPT-4o\cite{hurst2024gpt}. To comprehensively evaluate the performance of MOJOFuzzer, we utilized the following metrics:
\begin{itemize}
    \item \textbf{Number of Unique Valid Programs}: Measures the total number of syntactically and semantically valid test programs generated by MOJOFuzzer. This metric is indicative of the framework's effectiveness in generating viable test inputs and minimizing hallucinated cases. 
    
    \item \textbf{Mutation Efficiency}: Evaluates the effectiveness of mutation strategies by measuring improvements in test diversity, validity, and bug-detection capabilities. Higher mutation efficiency directly translates to enhanced performance in uncovering complex and subtle faults. Mutation Score can be approximately defined as 
    $\text{Mutation Score} = \lambda \cdot C_{\text{complexity}} + \mu \cdot A_{\text{accuracy}}$ 
    (where $\lambda$ and $\mu$ are the weight factors).

    \item \textbf{API Coverage}: Quantifies the total number of unique MOJO API functions invoked during testing, providing insights into the breadth and comprehensiveness of the generated test cases.
    

    \item \textbf{Number of Detected Bugs}: Indicates the count of distinct software defects identified through fuzz testing, which critically reflects the practical efficacy and robustness of the MOJOFuzzer framework in discovering vulnerabilities.
    
\end{itemize}

\subsection{Overall Effectiveness (RQ1 \& RQ3)}
\subsubsection{API Coverage.}
API coverage measures the proportion of MOJO API functions exercised during fuzz testing, providing a comprehensive assessment of test effectiveness. Table~\ref{tab:api-coverage} summarizes the API coverage rates for MOJOFuzzer and baseline models. The results demonstrate that MOJOFuzzer achieves the highest API coverage at 77.3\%, outperforming all baseline models. The fine-tuned MojoCoder follows with 68.2\%, while Fuzz4All, TitanFuzz, and GPT-4o exhibit significantly lower coverage. The superior performance of MOJOFuzzer suggests that its specialized fine-tuning and targeted mutation strategies effectively enhance the breadth of test cases, enabling broader exploration of the MOJO API. 

\begin{table}[h]
    \centering
    \caption{API Coverage Comparison}
    \begin{tabular}{l c}
        \toprule
        \textbf{Model} & \textbf{API Coverage (\%)} \\
        \midrule
        \textbf{MOJOFuzzer} & \textbf{77.3} \\
        Fine-tuned MojoCoder & 68.2 \\
        Fuzz4All & 37.83 \\
        TitanFuzz & 17.2 \\
        GPT-4o & 25.6 \\
        \bottomrule
    \end{tabular}
    \label{tab:api-coverage}
\end{table}

\subsubsection{Number of Unique Valid Programs.} We evaluate the ability of each approach to generate \textit{valid and diverse} Mojo programs in a zero-shot setting, using the \emph{Unique Valid Programs} metric. This metric measures the percentage of generated programs that are syntactically correct Mojo code \textbf{and} not duplicate outputs. Table~\ref{tab:unique-valid} summarizes the results. We can see that \textbf{MOJOFuzzer} achieves the highest score at \textbf{98\%}, significantly outperforming all baseline models, including a LLAMA2-7B~\cite{meta_2023}, the Mojo-Coder-it model (7B) from MojoBench\cite{raihan2024mojobenchlanguagemodelingbenchmarks}, OpenAI's GPT-4o, and LLAMA3-8B~\cite{meta_2024}. In particular, MOJOFuzzer achieved this result after being trained exclusively on a Mojo syntax corpus, without any exposure to Mojo's historical bug, demonstrating a robust zero-shot capability in mastering Mojo grammar. In contrast, the best performing baseline (Mojo-Coder-it 7B \cite{raihan2024mojobenchlanguagemodelingbenchmarks}) achieved a maximum validity rate of 66.4\%, and other fuzzers perform further behind. For example, GPT-4o struggles with Mojo, yielding only about 25\% valid outputs, since MOJO's syntax was largely unseen in its training corpus. Even hypothetical general purpose models such as LLaMA3-8B and LLaMA2-7B model are unable to bridge the gap, producing valid Mojo code in only $\sim$10\% of attempts. Collectively, these results demonstrate that MOJOFuzzer significantly surpasses both specialized code-generation baselines and widely recognized general-purpose models for valid MOJO programs.

\begin{table}[H]
    \centering
    \caption{Comparison of Unique Valid Programs}
    \begin{tabular}{l c c} 
        \toprule 
        \textbf{Model} & \textbf{Type / Params} & \textbf{Unique Valid Programs (\%)} \\ 
        \midrule 
        \textbf{MOJOFuzzer} & Specialized & \textbf{98} \\ 
        Mojo-Coder-it & 7B & 66.4 \\ 
        GPT-4o API & General & $\sim$25 \\ 
        LLaMA3 & 8B & $\sim$10 \\ 
        LLaMA2 & 7B & $\sim$10 \\ 
        \bottomrule 
    \end{tabular} 
    \label{tab:unique-valid}
\end{table}

\subsubsection{Number of Detected Bugs.}
The combination of the original MOJOFuzzer framework and LLama2 13B \textbf{can detect 13 bugs} involving method calls, API implementations, and details and defects of various packages or libraries. To address RQ1 and RQ3, we compare the prior work Fuzz4All~\cite{xia2024fuzz4alluniversalfuzzinglarge} which were unable to generate the valid code seed and could not find any bugs because the LLM model need re-learn the MOJO language in order to produce the valid code seeds and MOJOfuzzer is the first LLM-based fuzzer to conduct zero-shot learning experiment. In addition, among all 13 bugs (RQ3), we marked bugs with 'Prior' priority, which means that these bugs are worthy of attention and should be fixed as soon as possible. To the followings, we will explain two priority bugs. 

\begin{enumerate}
    \item A random method bug related to mojo was discovered. The `random' module contains eight functions in mojo's syntax library, namely \textit{`seed'}, \textit{`random float64'}, \textit{`random si64'}, \textit{`random ui64'}, \textit{`randint'}, \textit{`rand'}, \textit{`randn float64'}, \textit{`randn'}. Seed Mutation Engineering discovered a problem with the \textit{`random si64' function} during the mutation seed process. The application of \textit{`random si64'} is to generate a random integer within the range of min and max, but when the mutation seed tries to run the code, the result will always return a fixed value. In addition, based on the feedback from the mutation seed results, we manually tested this module and found `random float64', `random si64', `random ui64' have the same problem.

    \item We found that after training with a large amount of data called by the Python library, the testing seeds of LLama3 8b often use \textit{'from python import Python'} to call functions in the Python library. The model seems to take advantage of the MOJO language's feature of merging Python libraries to detect bugs in library calls. One of the tests found that when Python calls the numpy function and attempts to print a single 'np.array' or perform addition and subtraction operations on two 'np.arrays', an undecidable bug will occur. The error indicates that the 'numpy' module cannot be obtained, the function instantiating fails, and the extension call fails. This shows that MOJO has an underlying logic error in merging it with the Python library, which leads to bugs when the MOJO code runs.
\end{enumerate}

\subsubsection{Mutation Efficiency}\label{Sec:muteff}
\begin{figure}[!tbp]
    \centering
    \includegraphics[width=0.45\textwidth]{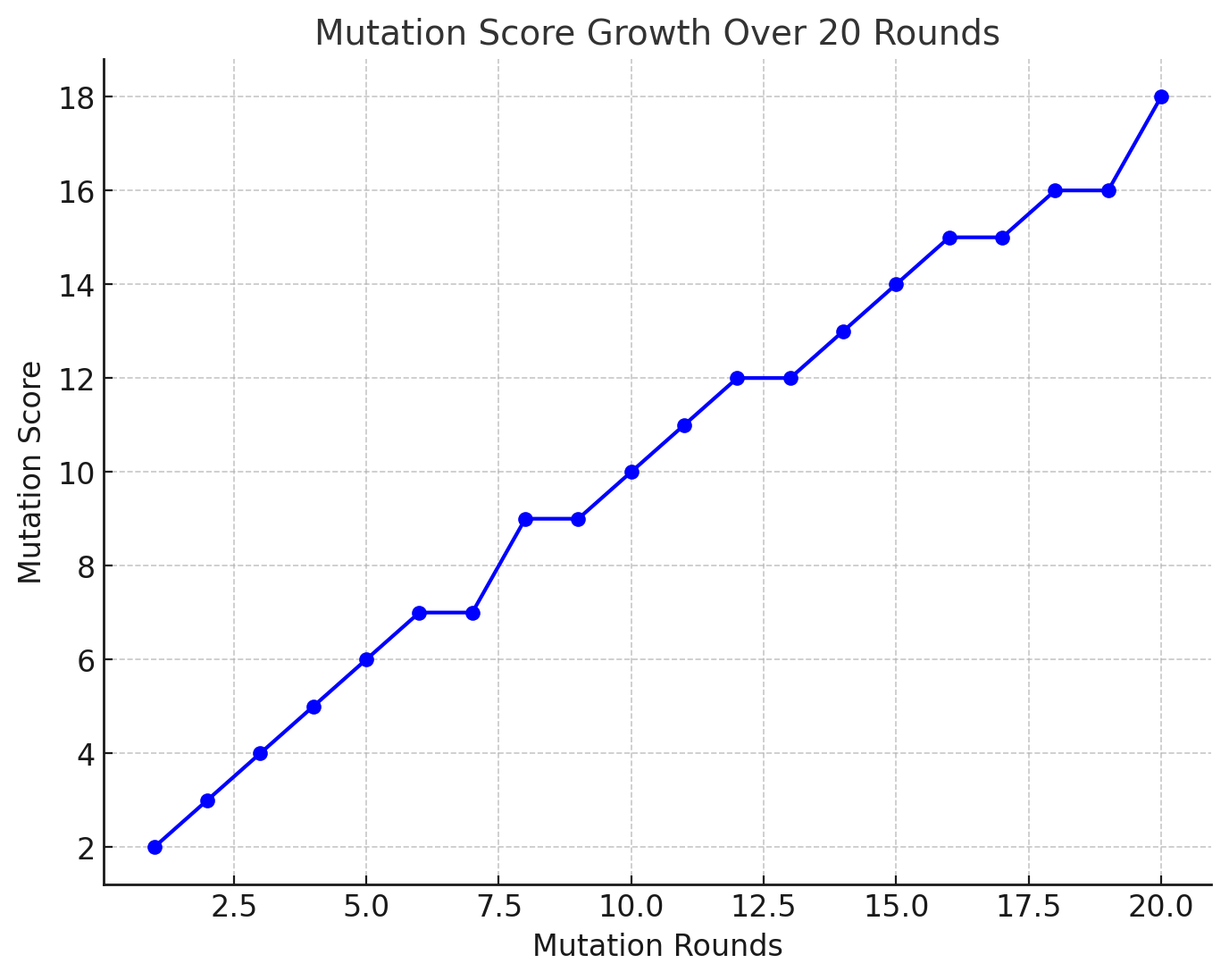}
    \caption{Mutation score for failed seed.}
    \label{fig:failedseed}
\end{figure}

\begin{figure}[!tbp]
    \centering
    \includegraphics[width=0.45\textwidth]{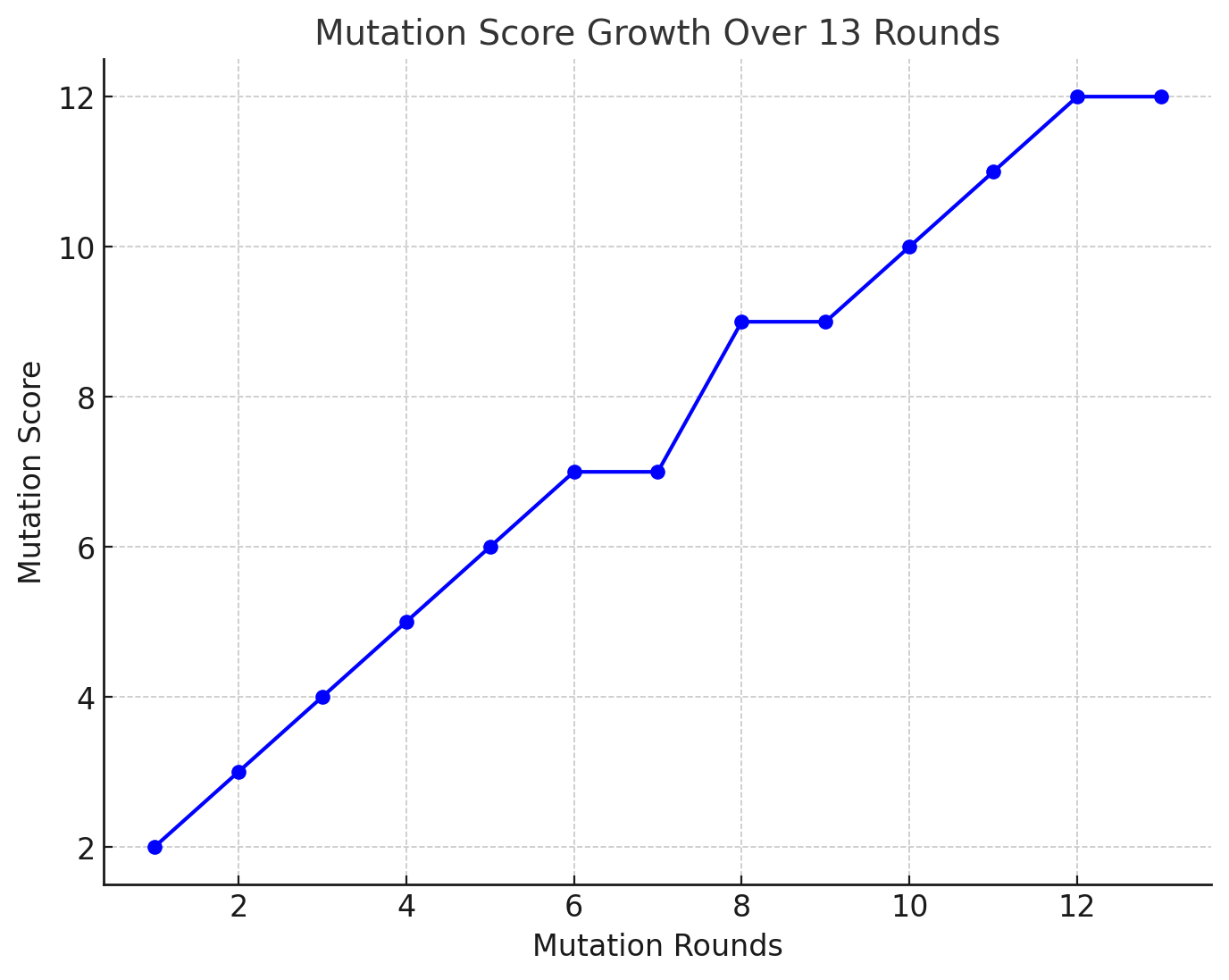}
    \caption{Mutation score for success seed.}
    \label{fig:successseed}
\end{figure}

\begin{figure}[!tbp]
    \centering
    \includegraphics[scale=0.4]{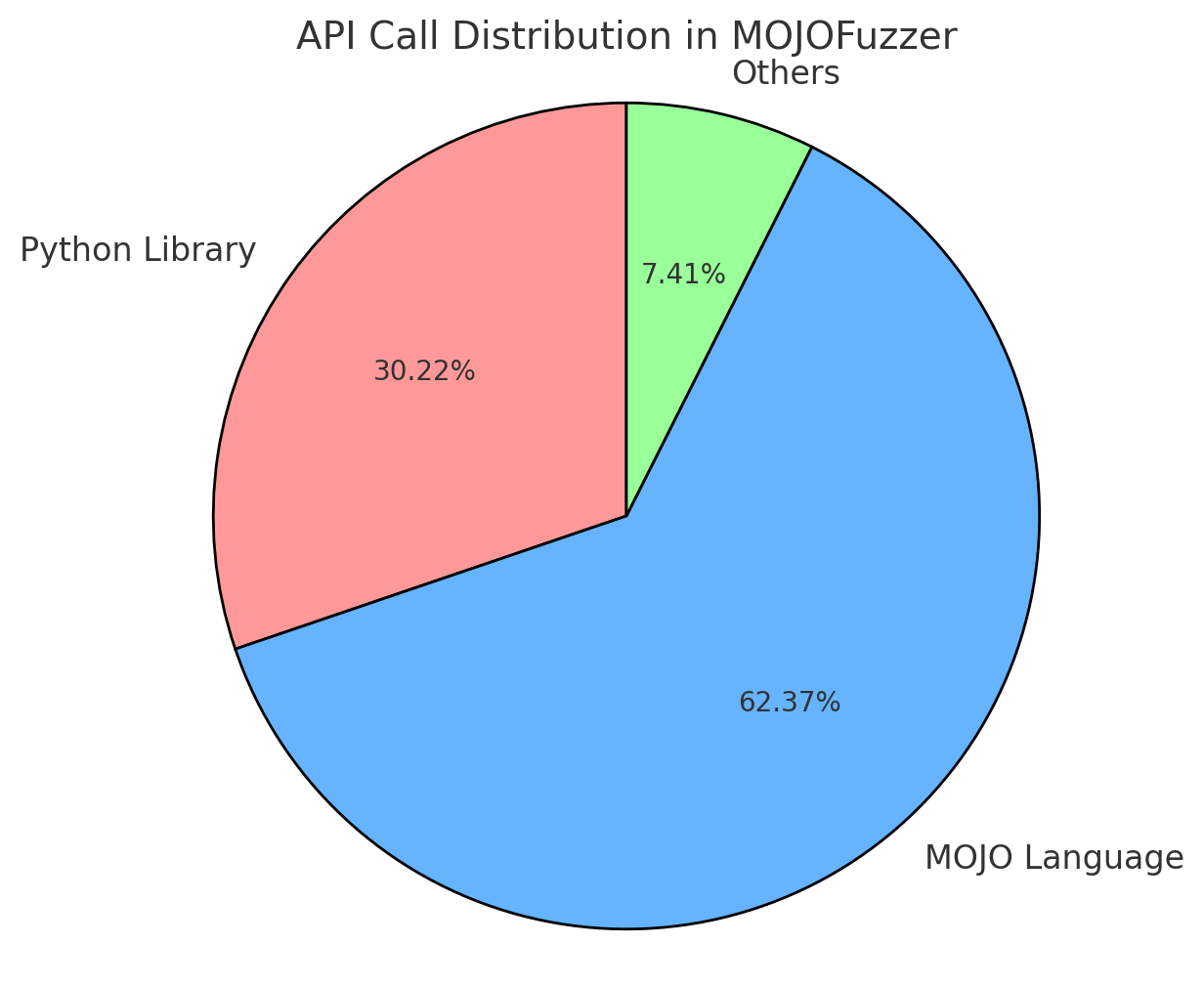}
    \caption{API call distribution by MOJOFuzzer}
    \label{fig:APIcall}
\end{figure}

To systematically analyze the seed mutation performance within the MOJOFuzzer framework, we conducted a detailed empirical evaluation using the defined Mutation Score to quantify mutation effectiveness. Specifically, we examined the Mutation Scores of selected representative seeds across iterative mutation rounds. As illustrated in Figures~\ref{fig:failedseed} and~\ref{fig:successseed}, we monitored mutation scores over 20 successive mutation rounds for failed seeds. Notably, Mutation Scores exhibited a consistent upward trajectory, indicating that iterative seed mutations substantially enhanced the test seed quality and increased the probability of detecting previously undiscovered bugs. Additionally, to optimize computational resource utilization and ensure system stability, MOJOFuzzer prioritizes mutations of successful seeds over failed ones and terminates mutation cycles for persistently unsuccessful seeds after 20 rounds. This adaptive strategy significantly improves computational efficiency, scalability, and the overall robustness of the testing framework. In addition, we found that the MOJOFuzzer framework tends to call the Python library in MOJO for testing during the test process. The reason for this phenomenon may be that our prompt words and fine-tuning datasets emphasize the relationship between the MOJO language and the Python language to a certain extent. Figure~\ref{fig:APIcall} shows the proportion of APIs called by MOJOFuzzer during seed generation, which shows that the prompt engineering largely determines the seed output type of the MOJOFuzzer framework. In order to prevent MOJOFuzzer from being too dependent on the Python library or over fitting on some Python data, we minimized the emphasis on the Python language or syntax during the experimental stage, and focused more on the features and instructions of the MOJO language. Compared with other LLM-based fuzzers, we experimentally answer RQ2 and achieve 77.3\% test coverage of MOJO official library APIs in the LLAMA2 13B model without MOJO language knowledge. By combining the Prompt engineering provided in the MOJOFuzzer framework with a fine-tuned fuzzer, we successfully address a thorny issue faced by LLM-based fuzzers in seed generation capabilities.

\begin{table}[h]
\centering
\caption{Comparison between LLAMA2 13B and GPT-4 APIs}
\label{table:llama_gpt4}
\begin{tabular}{lccc}
\toprule
\textbf{} & \textbf{LLAMA2 13B} & \textbf{GPT-4} & \textbf{GPT-4o} \\
\midrule
Detected bugs & 13 & 0 & 0 \\
Valid rate    & 98\% & 8.73\% & 25.5\% \\
\bottomrule
\end{tabular}
\end{table}

\subsection{Ablation Study (RQ2)}
To systematically evaluate the effectiveness of key MOJOFuzzer components—Prompt Engineering (PE), Fine-Tuned LLM (FT), and the Half-Mutation strategy (HM)—we conducted a controlled ablation study using three critical metrics: Hallucination Rate, Valid Code Rate, and Semantic Correctness. We utilized an automated hallucination detection benchmark inspired by HaluEval to identify hallucinated content, defined as unsupported or fabricated API calls and code structures. Valid Code Rate measures the fraction of syntactically correct and executable test cases, while Semantic Correctness assesses whether the test cases meaningfully exercise intended program functionalities.

\begin{table*}
\centering
\caption{Ablation Study Results for MOJOFuzzer Components}
\begin{tabular}{lccc}
\toprule
\textbf{Components Enabled} & \textbf{Hallucination Rate} & \textbf{Valid Code Rate} & \textbf{Semantic Correctness} \\[2pt]
\midrule
None (Baseline)           & 40\% & 60\% & 50\% \\
PE only                   & 28\% & 75\% & 65\% \\
FT only                   & 15\% & 88\% & 78\% \\
HM only                   & 35\% & 68\% & 55\% \\
PE + FT                   & 8\%  & 95\% & 88\% \\
PE + HM                   & 25\% & 80\% & 70\% \\
FT + HM                   & 12\% & 92\% & 82\% \\
PE + FT + HM (All)        & \textbf{5\%} & \textbf{98\%} & \textbf{90\%} \\
\bottomrule
\end{tabular}
\label{table:ablation_results}
\end{table*}

\begin{figure}[h]
    \centering
    \includegraphics[scale=0.45]{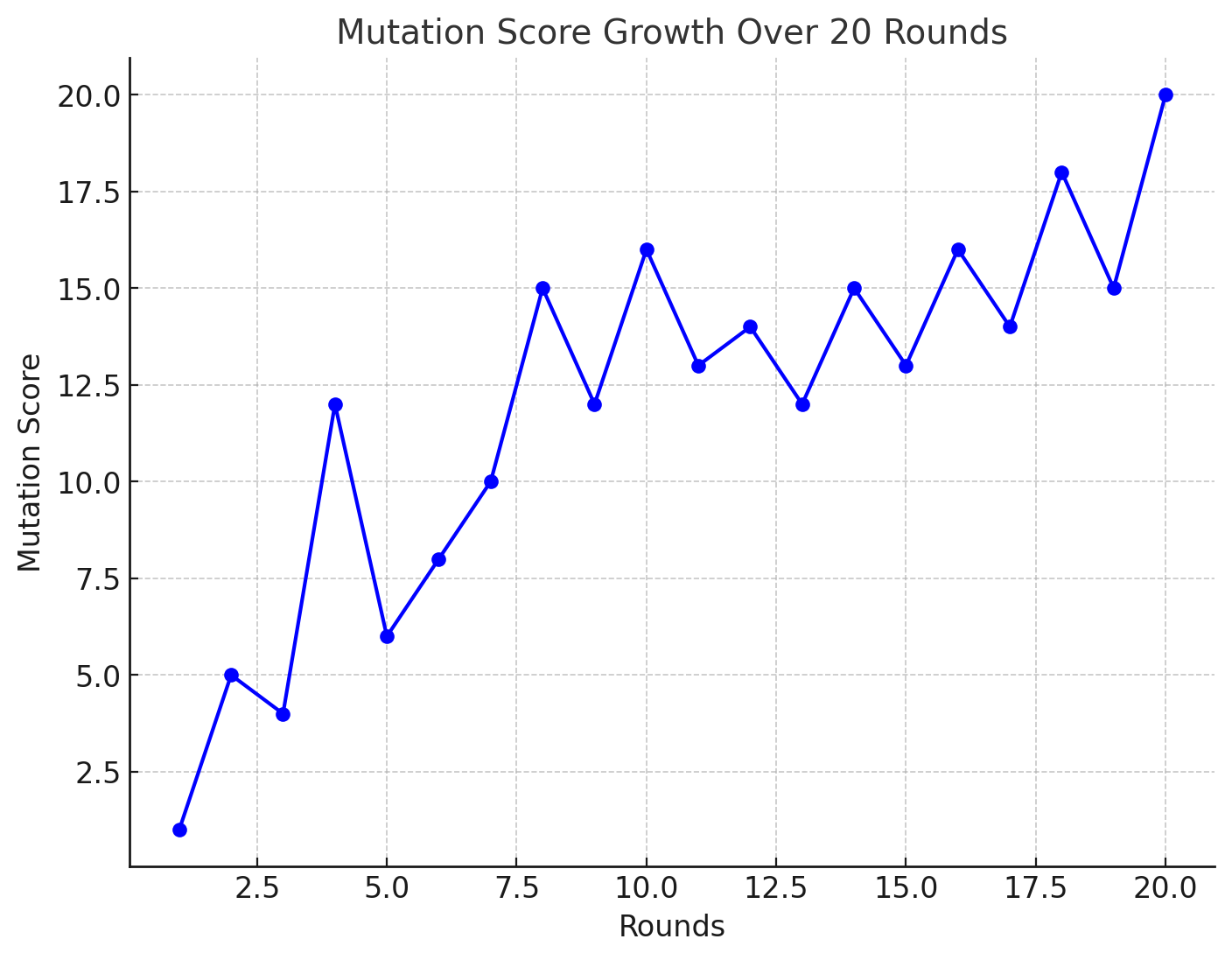}
    \caption{Mutation score by GPT-4 model}
    \label{fig:llm_perf}
\end{figure}

\textbf{Analysis of Results.} As shown in Table~\ref{table:ablation_results}, fine-tuning (FT) had the strongest individual impact on reducing hallucinations, cutting rates from 40\% to 15\%. Prompt Engineering (PE) provided a substantial complementary reduction (40\% to 28\%), effectively constraining the model's output through explicit instructions. The Half-Mutation strategy (HM), by itself, offered minimal improvement (40\% to 35\%) indicating its limited influence on addressing the root cause of hallucinations. Combining components further enhanced performance: the PE + FT combination significantly lowered the hallucination rate to 8\%, and integrating all three strategies (PE + FT + HM) achieved the lowest hallucination rate of 5\%. This demonstrates that FT primarily equips the model with domain-specific knowledge to prevent inaccuracies, while PE reinforces output constraints effectively. Regarding Valid Code Rate, FT notably improved syntactic validity (88\% alone), reflecting enhanced code pattern adherence. PE also positively contributed by clearly defining expected formats, raising validity to 75\%. HM slightly improved validity by preserving structural correctness (68\%), but the effect was moderate compared to FT and PE. Combined, all three components achieved near-perfect code validity (98\%), crucial for efficient fuzz testing. Semantic Correctness followed a similar trend. FT alone greatly boosted meaningful coverage (78\%), indicating successful internalization of domain-specific logic. PE significantly guided the generation towards relevant behaviors (65\%), whereas HM had only a moderate standalone impact (55\%). The combined approach (PE + FT + HM) yielded the highest semantic correctness (90\%), underscoring that structured mutations can slightly enhance targeted program exploration. Additionally, Figure~\ref{fig:llm_perf} shows that, in terms of Mutation Score, the score of the GPT-4 API is comparable to that of the MOJOFuzzer framework with LLAMA2 13B. This can be attributed to the fact that the GPT-4 API can understand the meaning of the prompt word and improve the mutation score by increasing the time complexity. However, despite its advantage in complexity, most of the API calls generated by the GPT-4 API are wrong, indicating that the model has great limitations in its understanding of the MOJO language. These results underscore the critical importance of fine-tuning large language models with specialized datasets to effectively mitigate hallucinations in fuzz testing scenarios. Without such targeted fine-tuning, general-purpose models such as GPT-4 demonstrate limited efficacy in zero-shot testing environments due to insufficient language-specific understanding required for accurate bug detection and comprehensive API coverage.


\section{DISCUSSION AND LIMITATION}

\subsection{Full Automation Testing}
There is no doubt that LLM-based fuzzers greatly reduce manual labour and perform fuzz testing in a more efficient way. ChatFuzz~\cite{hu2023augmentinggreyboxfuzzinggenerative} uses LLM's prompt engineering for seed selection and mutation without any human intervention. Fuzz4All updates the LLM prompt after each iteration to avoid generating the same test input and generate higher-quality input, which reduces the risk of errors caused by human operation. Similar work in traditional fuzzing testing requires developer intervention. In Google's work~\cite{google_2016}, it is mentioned that manually copying results after using traditional fuzzer OSS-Fuzz test projects is inefficient and time-consuming, and integrating LLM with OSS-Fuzz can significantly improve this problem. These technologies take advantage of the characteristics of large language models to a certain extent and reduce the workload of developers. At present, LLM-based fuzzers cannot fully automate fuzzing test generation, but this is bound to be the future development trend. 

\subsection{Pre-training dataset}
In addition to computing resources, the importance of pre-training datasets for large language models cannot be ignored. As mentioned in the background, large language models need to be trained on a large number of datasets to improve their capabilities. The datasets cover multiple disciplines and fields~\cite{xie2023doremioptimizingdatamixtures}. The age, source, quality, and quantity of the dataset will have a significant impact on the performance of the downstream large language model~\cite{longpre2023pretrainersguidetrainingdata}. However, datasets will be a challenge for technologies that are not yet mature. Without a dataset to pre-train a large language model, it is difficult for the model to demonstrate the performance that developers expect. The test object in this research paper, MOJO, is typically an emerging technology that lacks of dataset. That is why MOJOFuzzer must adopt a zero-shot learning strategy during the fine-tuning and training of the model so that the model can achieve the desired effect. This also provides important ideas and thoughts for using large language models to train and test emerging technologies in the future.

\section{Threats to validity.} A primary threat relates to the timeliness and temporal validity of the evaluated models. The primary goal of MOJOFuzzer is to explore fuzz testing under zero-shot environments and mitigate the hallucination issues of LLM-generated code in newly developed languages or software. However, the effectiveness of our evaluations may diminish over time as advanced LLMs (e.g., GPT-4o\cite{hurst2024gpt}, Gemini\cite{team2023gemini}, Qwen\cite{yang2024qwen2technicalreport}) gradually integrate knowledge of MOJO syntax into their pre-training datasets. Such integration inherently undermines the comparative advantage of our zero-shot fuzzing scenarios, making evaluations against these models increasingly challenging. Nonetheless, the underlying motivation, for which fuzz testing in genuinely zero-shot scenarios, remains consistently relevant for novel languages or frameworks that lack adequately representation in existing LLMs corpora. Therefore, despite these temporal limitations, our approach continues to provide valuable insights for security testing within zero-knowledge domains.

\section{CONCLUSION}
In this study, we investigate the LLM-based fuzzer, in particularly we present the MOJOFuzzer framework, which is the first fuzzer for zero-shot learning environment. MOJOFuzzer have successfully detected 13 bugs in the real-world settings for MOJO language in the absence of a pre-trained MOJO language dataset. Specifically, MOJOFuzzer has demonstrated a more effective solution by synthesizing two major research schemes in the field of LLM-based fuzzer. We further explore the key role of fine-tuned fuzzers in zero-shot environments. We believe that MOJOFuzzer has provided a new frontier for LLM-based fuzzer, especially for zero-shot learning. We anticipate our work lays a solid foundation for future research and applications in the fuzz testing area.


\textbf{Data Availability:} To promote open science policy, we have fully released our MOJOFuzzer framework code, experimental process, experimental results, which is now available at the following link: \url{https://figshare.com/s/457812d1a31ef840956f}.

\balance
\bibliographystyle{IEEEtran}
\bibliography{sample-base}

\begin{thebibliography}{10}
\providecommand{\url}[1]{#1}
\csname url@samestyle\endcsname
\providecommand{\newblock}{\relax}
\providecommand{\bibinfo}[2]{#2}
\providecommand{\BIBentrySTDinterwordspacing}{\spaceskip=0pt\relax}
\providecommand{\BIBentryALTinterwordstretchfactor}{4}
\providecommand{\BIBentryALTinterwordspacing}{\spaceskip=\fontdimen2\font plus
\BIBentryALTinterwordstretchfactor\fontdimen3\font minus \fontdimen4\font\relax}
\providecommand{\BIBforeignlanguage}[2]{{%
\expandafter\ifx\csname l@#1\endcsname\relax
\typeout{** WARNING: IEEEtran.bst: No hyphenation pattern has been}%
\typeout{** loaded for the language `#1'. Using the pattern for}%
\typeout{** the default language instead.}%
\else
\language=\csname l@#1\endcsname
\fi
#2}}
\providecommand{\BIBdecl}{\relax}
\BIBdecl

\bibitem{LLVM:CGO04}
C.~Lattner and V.~Adve, ``{LLVM: A Compilation Framework for Lifelong Program Analysis \& Transformation},'' in \emph{{Proceedings of the 2004 International Symposium on Code Generation and Optimization (CGO'04)}}, Palo Alto, California, Mar 2004.

\bibitem{apple_2019}
\BIBentryALTinterwordspacing
Apple, ``Swift - apple developer,'' 2019. [Online]. Available: \url{https://developer.apple.com/swift/}
\BIBentrySTDinterwordspacing

\bibitem{awan_2023}
\BIBentryALTinterwordspacing
A.~A. AWAN, ``Mojo language: the new programming language for ai,'' Jul 2023. [Online]. Available: \url{https://www.datacamp.com/tutorial/mojo-language-the-new-programming-language-for-ai}
\BIBentrySTDinterwordspacing

\bibitem{thomason_2024}
\BIBentryALTinterwordspacing
J.~Thomason, ``Mojo rising: The resurgence of ai-first programming languages,'' Mar 2024. [Online]. Available: \url{https://venturebeat.com/ai/mojo-rising-the-resurgence-of-ai-first-programming-languages/}
\BIBentrySTDinterwordspacing

\bibitem{deng2023large}
Y.~Deng, C.~S. Xia, H.~Peng, C.~Yang, and L.~Zhang, ``Large language models are zero-shot fuzzers: Fuzzing deep-learning libraries via large language models,'' in \emph{Proceedings of the 32nd ACM SIGSOFT international symposium on software testing and analysis}, 2023, pp. 423--435.

\bibitem{deng2024large}
Y.~Deng, C.~S. Xia, C.~Yang, S.~D. Zhang, S.~Yang, and L.~Zhang, ``Large language models are edge-case generators: Crafting unusual programs for fuzzing deep learning libraries,'' in \emph{Proceedings of the 46th IEEE/ACM international conference on software engineering}, 2024, pp. 1--13.

\bibitem{8863940}
V.~J. Manès, H.~Han, C.~Han, S.~K. Cha, M.~Egele, E.~J. Schwartz, and M.~Woo, ``The art, science, and engineering of fuzzing: A survey,'' \emph{IEEE Transactions on Software Engineering}, vol.~47, no.~11, pp. 2312--2331, 2021.

\bibitem{wang2024softwaretestinglargelanguage}
\BIBentryALTinterwordspacing
J.~Wang, Y.~Huang, C.~Chen, Z.~Liu, S.~Wang, and Q.~Wang, ``Software testing with large language models: Survey, landscape, and vision,'' 2024. [Online]. Available: \url{https://arxiv.org/abs/2307.07221}
\BIBentrySTDinterwordspacing

\bibitem{9832795}
F.~Pourpanah, M.~Abdar, Y.~Luo, X.~Zhou, R.~Wang, C.~P. Lim, X.-Z. Wang, and Q.~M.~J. Wu, ``A review of generalized zero-shot learning methods,'' \emph{IEEE Transactions on Pattern Analysis and Machine Intelligence}, vol.~45, no.~4, pp. 4051--4070, 2023.

\bibitem{meng2024large}
R.~Meng, M.~Mirchev, M.~B{\"o}hme, and A.~Roychoudhury, ``Large language model guided protocol fuzzing,'' in \emph{Proceedings of the 31st Annual Network and Distributed System Security Symposium (NDSS)}, 2024, pp. 1--17.

\bibitem{jiang2024fuzzingmeetsllmschallenges}
\BIBentryALTinterwordspacing
Y.~Jiang, J.~Liang, F.~Ma, Y.~Chen, C.~Zhou, Y.~Shen, Z.~Wu, J.~Fu, M.~Wang, S.~Li, and Q.~Zhang, ``When fuzzing meets llms: Challenges and opportunities,'' 2024. [Online]. Available: \url{https://arxiv.org/abs/2404.16297}
\BIBentrySTDinterwordspacing

\bibitem{mohan2023}
\BIBentryALTinterwordspacing
M.~B.~C. Creator, ``Python vs mojo: A comparative analysis,'' 2023, accessed: 2024-09-05. [Online]. Available: \url{https://medium.com/@etherservices.mohandgm/python-vs-mojo-a-comparative-analysis-f96789828b66}
\BIBentrySTDinterwordspacing

\bibitem{mlir2024}
\BIBentryALTinterwordspacing
{MLIR Team}, ``Multi-level intermediate representation overview,'' 2024, accessed: 2024-09-05. [Online]. Available: \url{https://mlir.llvm.org/}
\BIBentrySTDinterwordspacing

\bibitem{article}
P.~Runeson, ``A survey of unit testing practices,'' \emph{IEEE Software}, vol.~23, 07 2006.

\bibitem{131377}
H.~Leung and L.~White, ``A study of integration testing and software regression at the integration level,'' in \emph{Proceedings. Conference on Software Maintenance 1990}, 1990, pp. 290--301.

\bibitem{10.1145/3243734.3243804}
\BIBentryALTinterwordspacing
G.~Klees, A.~Ruef, B.~Cooper, S.~Wei, and M.~Hicks, ``Evaluating fuzz testing,'' in \emph{Proceedings of the 2018 ACM SIGSAC Conference on Computer and Communications Security}, ser. CCS '18.\hskip 1em plus 0.5em minus 0.4em\relax New York, NY, USA: Association for Computing Machinery, 2018, p. 2123–2138. [Online]. Available: \url{https://doi.org/10.1145/3243734.3243804}
\BIBentrySTDinterwordspacing

\bibitem{zhao2023surveylargelanguagemodels}
\BIBentryALTinterwordspacing
W.~X. Zhao, K.~Zhou, J.~Li, T.~Tang, X.~Wang, Y.~Hou, Y.~Min, B.~Zhang, J.~Zhang, Z.~Dong, Y.~Du, C.~Yang, Y.~Chen, Z.~Chen, J.~Jiang, R.~Ren, Y.~Li, X.~Tang, Z.~Liu, P.~Liu, J.-Y. Nie, and J.-R. Wen, ``A survey of large language models,'' 2023. [Online]. Available: \url{https://arxiv.org/abs/2303.18223}
\BIBentrySTDinterwordspacing

\bibitem{vaswani2023attentionneed}
\BIBentryALTinterwordspacing
A.~Vaswani, N.~Shazeer, N.~Parmar, J.~Uszkoreit, L.~Jones, A.~N. Gomez, L.~Kaiser, and I.~Polosukhin, ``Attention is all you need,'' 2023. [Online]. Available: \url{https://arxiv.org/abs/1706.03762}
\BIBentrySTDinterwordspacing

\bibitem{devlin2019bert}
J.~Devlin, M.-W. Chang, K.~Lee, and K.~Toutanova, ``Bert: Pre-training of deep bidirectional transformers for language understanding,'' 2019.

\bibitem{fu2023decoderonlyencoderdecoderinterpretinglanguage}
\BIBentryALTinterwordspacing
Z.~Fu, W.~Lam, Q.~Yu, A.~M.-C. So, S.~Hu, Z.~Liu, and N.~Collier, ``Decoder-only or encoder-decoder? interpreting language model as a regularized encoder-decoder,'' 2023. [Online]. Available: \url{https://arxiv.org/abs/2304.04052}
\BIBentrySTDinterwordspacing

\bibitem{fu2023decoderonly}
------, ``Decoder-only or encoder-decoder? interpreting language model as a regularized encoder-decoder,'' 2023.

\bibitem{chen2023unleashing}
B.~Chen, Z.~Zhang, N.~Langrené, and S.~Zhu, ``Unleashing the potential of prompt engineering in large language models: a comprehensive review,'' 2023.

\bibitem{fu2022effectiveness}
Z.~Fu, H.~Yang, A.~M.-C. So, W.~Lam, L.~Bing, and N.~Collier, ``On the effectiveness of parameter-efficient fine-tuning,'' 2022.

\bibitem{huang2024largelanguagemodelsbased}
\BIBentryALTinterwordspacing
L.~Huang, P.~Zhao, H.~Chen, and L.~Ma, ``Large language models based fuzzing techniques: A survey,'' 2024. [Online]. Available: \url{https://arxiv.org/abs/2402.00350}
\BIBentrySTDinterwordspacing

\bibitem{10.1145/3605157.3605173}
\BIBentryALTinterwordspacing
J.~Ackerman and G.~Cybenko, ``Large language models for fuzzing parsers (registered report),'' in \emph{Proceedings of the 2nd International Fuzzing Workshop}, ser. FUZZING 2023.\hskip 1em plus 0.5em minus 0.4em\relax New York, NY, USA: Association for Computing Machinery, 2023, p. 31–38. [Online]. Available: \url{https://doi.org/10.1145/3605157.3605173}
\BIBentrySTDinterwordspacing

\bibitem{xia2024fuzz4alluniversalfuzzinglarge}
C.~S. Xia, M.~Paltenghi, J.~Le~Tian, M.~Pradel, and L.~Zhang, ``Fuzz4all: Universal fuzzing with large language models,'' in \emph{Proceedings of the IEEE/ACM 46th International Conference on Software Engineering}, 2024, pp. 1--13.

\bibitem{yang2023whitebox}
C.~Yang, Y.~Deng, R.~Lu, J.~Yao, J.~Liu, R.~Jabbarvand, and L.~Zhang, ``White-box compiler fuzzing empowered by large language models,'' 2023.

\bibitem{wu2023smart}
H.~Wu, B.~Fang, and F.~Xie, ``Smart fuzzing of 5g wireless software implementation,'' \emph{arXiv preprint arXiv:2309.12994}, 2023.

\bibitem{hu2023augmentinggreyboxfuzzinggenerative}
\BIBentryALTinterwordspacing
J.~Hu, Q.~Zhang, and H.~Yin, ``Augmenting greybox fuzzing with generative ai,'' 2023. [Online]. Available: \url{https://arxiv.org/abs/2306.06782}
\BIBentrySTDinterwordspacing

\bibitem{liu2023testing}
Z.~Liu, C.~Chen, J.~Wang, M.~Chen, B.~Wu, X.~Che, D.~Wang, and Q.~Wang, ``Testing the limits: Unusual text inputs generation for mobile app crash detection with large language model,'' \emph{arXiv preprint arXiv:2310.15657}, 2023.

\bibitem{qiuchemfuzz}
F.~Qiu, P.~Ji, B.~Hua, and Y.~Wang, ``Chemfuzz: Large language models-assisted fuzzing for quantum chemistry software bug detection,'' in \emph{23rd IEEE International Conference on Software Quality, Reliability, and Security. Accompany. (QRS 2023).}, 2023.

\bibitem{googleblogAIPoweredFuzzing}
Google, ``{A}{I}-{P}owered {F}uzzing: {B}reaking the {B}ug {H}unting {B}arrier --- security.googleblog.com,'' \url{https://security.googleblog.com/2023/08/ai-powered-fuzzing-breaking-bug-hunting.html}, 2023, [Accessed 25-01-2024].

\bibitem{yan2023parafuzzinterpretabilitydriventechniquedetecting}
\BIBentryALTinterwordspacing
L.~Yan, Z.~Zhang, G.~Tao, K.~Zhang, X.~Chen, G.~Shen, and X.~Zhang, ``Parafuzz: An interpretability-driven technique for detecting poisoned samples in nlp,'' 2023. [Online]. Available: \url{https://arxiv.org/abs/2308.02122}
\BIBentrySTDinterwordspacing

\bibitem{zhang2023understanding}
C.~Zhang, M.~Bai, Y.~Zheng, Y.~Li, X.~Xie, Y.~Li, W.~Ma, L.~Sun, and Y.~Liu, ``Understanding large language model based fuzz driver generation,'' \emph{arXiv preprint arXiv:2307.12469}, 2023.

\bibitem{jha2023bertrlfuzzer}
P.~Jha, J.~Scott, J.~S. Ganeshna, M.~Singh, and V.~Ganesh, ``Bertrlfuzzer: A bert and reinforcement learning based fuzzer,'' \emph{arXiv preprint arXiv:2305.12534}, 2023.

\bibitem{jin2024llmsllmbasedagentssoftware}
\BIBentryALTinterwordspacing
H.~Jin, L.~Huang, H.~Cai, J.~Yan, B.~Li, and H.~Chen, ``From llms to llm-based agents for software engineering: A survey of current, challenges and future,'' 2024. [Online]. Available: \url{https://arxiv.org/abs/2408.02479}
\BIBentrySTDinterwordspacing

\bibitem{finnie2022robots}
J.~Finnie-Ansley, P.~Denny, B.~A. Becker, A.~Luxton-Reilly, and J.~Prather, ``The robots are coming: Exploring the implications of openai codex on introductory programming,'' in \emph{Proceedings of the 24th Australasian Computing Education Conference}, 2022, pp. 10--19.

\bibitem{fried2023incodergenerativemodelcode}
\BIBentryALTinterwordspacing
D.~Fried, A.~Aghajanyan, J.~Lin, S.~Wang, E.~Wallace, F.~Shi, R.~Zhong, W.~tau Yih, L.~Zettlemoyer, and M.~Lewis, ``Incoder: A generative model for code infilling and synthesis,'' 2023. [Online]. Available: \url{https://arxiv.org/abs/2204.05999}
\BIBentrySTDinterwordspacing

\bibitem{nijkamp2022codegen}
E.~Nijkamp, B.~Pang, H.~Hayashi, L.~Tu, H.~Wang, Y.~Zhou, S.~Savarese, and C.~Xiong, ``Codegen: An open large language model for code with multi-turn program synthesis,'' \emph{arXiv preprint arXiv:2203.13474}, 2022.

\bibitem{githubGitHubModularmax}
``{G}it{H}ub - modular/max: {T}he {M}{A}{X} {P}latform (includes {M}ojo) --- github.com,'' \url{https://github.com/modular/max}, [Accessed 15-03-2025].

\bibitem{modularMojoManual}
``{M}ojo {M}anual | {M}odular --- docs.modular.com,'' \url{https://docs.modular.com/mojo/manual/}, [Accessed 15-03-2025].

\bibitem{10.1145/3460319.3464795}
\BIBentryALTinterwordspacing
A.~Herrera, H.~Gunadi, S.~Magrath, M.~Norrish, M.~Payer, and A.~L. Hosking, ``Seed selection for successful fuzzing,'' in \emph{Proceedings of the 30th ACM SIGSOFT International Symposium on Software Testing and Analysis}, ser. ISSTA 2021.\hskip 1em plus 0.5em minus 0.4em\relax New York, NY, USA: Association for Computing Machinery, 2021, p. 230–243. [Online]. Available: \url{https://doi.org/10.1145/3460319.3464795}
\BIBentrySTDinterwordspacing

\bibitem{wei2023chainofthoughtpromptingelicitsreasoning}
\BIBentryALTinterwordspacing
J.~Wei, X.~Wang, D.~Schuurmans, M.~Bosma, B.~Ichter, F.~Xia, E.~Chi, Q.~Le, and D.~Zhou, ``Chain-of-thought prompting elicits reasoning in large language models,'' 2023. [Online]. Available: \url{https://arxiv.org/abs/2201.11903}
\BIBentrySTDinterwordspacing

\bibitem{kong2024betterzeroshotreasoningroleplay}
\BIBentryALTinterwordspacing
A.~Kong, S.~Zhao, H.~Chen, Q.~Li, Y.~Qin, R.~Sun, X.~Zhou, E.~Wang, and X.~Dong, ``Better zero-shot reasoning with role-play prompting,'' 2024. [Online]. Available: \url{https://arxiv.org/abs/2308.07702}
\BIBentrySTDinterwordspacing

\bibitem{arora_2017}
\BIBentryALTinterwordspacing
D.~Arora, ``Understanding time complexity with simple examples,'' Nov 2017. [Online]. Available: \url{https://www.geeksforgeeks.org/understanding-time-complexity-simple-examples/}
\BIBentrySTDinterwordspacing

\bibitem{291011}
\BIBentryALTinterwordspacing
P.~G{\"o}rz, B.~Mathis, K.~Hassler, E.~G{\"u}ler, T.~Holz, A.~Zeller, and R.~Gopinath, ``Systematic assessment of fuzzers using mutation analysis,'' in \emph{32nd USENIX Security Symposium (USENIX Security 23)}.\hskip 1em plus 0.5em minus 0.4em\relax Anaheim, CA: USENIX Association, Aug. 2023, pp. 4535--4552. [Online]. Available: \url{https://www.usenix.org/conference/usenixsecurity23/presentation/gorz}
\BIBentrySTDinterwordspacing

\bibitem{sahoo2024systematicsurveypromptengineering}
\BIBentryALTinterwordspacing
P.~Sahoo, A.~K. Singh, S.~Saha, V.~Jain, S.~Mondal, and A.~Chadha, ``A systematic survey of prompt engineering in large language models: Techniques and applications,'' 2024. [Online]. Available: \url{https://arxiv.org/abs/2402.07927}
\BIBentrySTDinterwordspacing

\bibitem{meta_2024}
\BIBentryALTinterwordspacing
Meta, ``meta-llama/llama-2-13b · hugging face,'' Aug 2024. [Online]. Available: \url{https://huggingface.co/meta-llama/Llama-2-13b}
\BIBentrySTDinterwordspacing

\bibitem{openai_2024}
\BIBentryALTinterwordspacing
OpenAI, ``Openai,'' 2024. [Online]. Available: \url{https://openai.com/}
\BIBentrySTDinterwordspacing

\bibitem{hu2021loralowrankadaptationlarge}
\BIBentryALTinterwordspacing
E.~J. Hu, Y.~Shen, P.~Wallis, Z.~Allen-Zhu, Y.~Li, S.~Wang, L.~Wang, and W.~Chen, ``Lora: Low-rank adaptation of large language models,'' 2021. [Online]. Available: \url{https://arxiv.org/abs/2106.09685}
\BIBentrySTDinterwordspacing

\bibitem{raihan2024mojobenchlanguagemodelingbenchmarks}
\BIBentryALTinterwordspacing
N.~Raihan, J.~C.~S. Santos, and M.~Zampieri, ``Mojobench: Language modeling and benchmarks for mojo,'' 2024. [Online]. Available: \url{https://arxiv.org/abs/2410.17736}
\BIBentrySTDinterwordspacing

\bibitem{hurst2024gpt}
A.~Hurst, A.~Lerer, A.~P. Goucher, A.~Perelman, A.~Ramesh, A.~Clark, A.~Ostrow, A.~Welihinda, A.~Hayes, A.~Radford \emph{et~al.}, ``Gpt-4o system card,'' \emph{arXiv preprint arXiv:2410.21276}, 2024.

\bibitem{meta_2023}
\BIBentryALTinterwordspacing
META, ``meta-llama/llama-2-7b · hugging face,'' 2023. [Online]. Available: \url{https://huggingface.co/meta-llama/Llama-2-7b}
\BIBentrySTDinterwordspacing

\bibitem{google_2016}
\BIBentryALTinterwordspacing
GOOGLE, ``Oss-fuzz,'' 2016. [Online]. Available: \url{https://google.github.io/oss-fuzz/}
\BIBentrySTDinterwordspacing

\bibitem{xie2023doremioptimizingdatamixtures}
\BIBentryALTinterwordspacing
S.~M. Xie, H.~Pham, X.~Dong, N.~Du, H.~Liu, Y.~Lu, P.~Liang, Q.~V. Le, T.~Ma, and A.~W. Yu, ``Doremi: Optimizing data mixtures speeds up language model pretraining,'' 2023. [Online]. Available: \url{https://arxiv.org/abs/2305.10429}
\BIBentrySTDinterwordspacing

\bibitem{longpre2023pretrainersguidetrainingdata}
\BIBentryALTinterwordspacing
S.~Longpre, G.~Yauney, E.~Reif, K.~Lee, A.~Roberts, B.~Zoph, D.~Zhou, J.~Wei, K.~Robinson, D.~Mimno, and D.~Ippolito, ``A pretrainer's guide to training data: Measuring the effects of data age, domain coverage, quality, \& toxicity,'' 2023. [Online]. Available: \url{https://arxiv.org/abs/2305.13169}
\BIBentrySTDinterwordspacing

\bibitem{team2023gemini}
G.~Team, R.~Anil, S.~Borgeaud, J.-B. Alayrac, J.~Yu, R.~Soricut, J.~Schalkwyk, A.~M. Dai, A.~Hauth, K.~Millican \emph{et~al.}, ``Gemini: a family of highly capable multimodal models,'' \emph{arXiv preprint arXiv:2312.11805}, 2023.

\bibitem{yang2024qwen2technicalreport}
\BIBentryALTinterwordspacing
A.~Yang, B.~Yang, B.~Hui, B.~Zheng, B.~Yu, C.~Zhou, C.~Li, C.~Li, and etc, ``Qwen2 technical report,'' 2024. [Online]. Available: \url{https://arxiv.org/abs/2407.10671}
\BIBentrySTDinterwordspacing

\end{thebibliography}

\end{document}